\documentstyle[preprint,prb,aps,epsf]{revtex}

\tightenlines
\begin{document}
\draft 

\title{Structure Space of Model Proteins:\\
A Principal Component Analysis}
\author{Mehdi Yahyanejad, Mehran Kardar}
\address{Department of Physics, Massachussetts Institute of Technology,
Cambridge, Massachusetts 02139}

\author{Chao Tang\thanks{Author to whom correspondence should be
addressed; electronic mail: tang@research.nj.nec.com}}
\address{NEC Research Institute, 4 Independence Way, Princeton, New
Jersey 08540}

\maketitle

\begin{abstract}
We study the space of all compact structures on a two-dimensional
square lattice of size $N=6\times6$. Each structure is mapped onto a
vector in $N$-dimensions according to a hydrophobic model. Previous
work has shown that the designabilities of structures are closely
related to the distribution of the structure vectors in the $N$-dimensional
space, with highly designable structures predominantly found in
low density regions. 
We use principal component analysis to probe and characterize the
distribution of structure vectors, and find a non-uniform density with a single peak.
Interestingly,  the principal axes of this peak are almost aligned
with Fourier eigenvectors, and the corresponding Fourier eigenvalues
go to zero continuously at the wave-number for alternating patterns ($q=\pi$).
These observations provide a stepping stone for an analytic description
of the distribution of structural points, and open the possibility of
estimating designabilities of realistic structures by simply Fourier
transforming the hydrophobicities of the corresponding sequences.

\end{abstract}

\newpage
\narrowtext

\section{INTRODUCTION}

Proteins fold into specific structures to perform their biological
function. Despite the huge diversity in their functions, evolutional
paths, structural details, and sequences, the vast majority of proteins
adopt only a small number ($\sim 1000$) of folds (``topology''). 
\cite{Finkel87,Chothia92,Brenner97,Orengo94,Wang96,Govin99}
This observation has intrigued a number of authors and lead to the concept
of {\em designability}. \cite{Finkel87,Camacho93,Yue95,Govin95,Li96}
The designability of a structure is defined to be the number of sequences 
that have that structure as their unique lowest-energy state. \cite{Li96}
It has been shown in various model studies that structures differ
drastically in their designability; a small number of highly designable
structures emerge with their associated number of sequences much larger
than the average. \cite{Li96,Govin96,Li98,Buchler00,Helling01,Cejtin02,Miller02}
Highly designable structures are also found to possess other protein-like
properties, such as thermodynamic stability, \cite{Li96} fast folding
kinetics, \cite{Govin95,Melin99} and tertiary symmetry. 
\cite{Yue95,Li96,Wang00}
These results suggest that there may be a designability principal behind
nature's selection of protein folds; these small number of folds were
selected because they are readily designed, stable against mutations,
and thermodynamically stable.

Why are some structures more designable than others? How do we identify
highly designable structures? Finkelstein and co-workers 
argued that certain motifs are easier to
stabilize and thus more common because they either have lower (e.g.
bending) energies or have unusual energy spectra over random sequences.
\cite{Finkel87,Finkel93,Finkel95} 
Govindarajan and Goldstein suggested
that the topology of a protein structure should be such that it is
kinetically ``foldable''. \cite{Govin95,Govin96,Govin97} More recently,
it was noted that an important clue resides  in the distribution of 
structures in a suitably defined structure space, with highly designable structures 
located in regions of low density. \cite{Li98,Buchler00} In particular, within a 
hydrophobic model, Li {\em et al.} showed that the distribution of 
structures is very nonuniform, and that the highly designable structures
are those that are far away from other structures. \cite{Li98} However, 
identifying highly designable structures still remains a tedious task,
requiring either full enumeration or sufficiently large sampling of both
the structure and the sequence spaces, making studies of large systems
prohibitive.

In this paper, we investigate the properties of the structure space of
the hydrophobic model of Li {\em et al.}, starting from a 
Principal Component  Analysis (PCA).  
We show that while the distribution of the structures is not uniform, 
it can be approximated as a cloud of points centered on a single peak.
The principal directions of this cloud are almost coincident with those
obtained by rotation into Fourier space; the coincidence is in fact exact
for the subset of cyclic structures.
An interesting feature is that the eigenvalues of PCA, describing the
extent of the density cloud along the principal axis, vary continuously
with the Fourier label $q$, with a minimum at $q=\pi$ corresponding
to alternating patterns.
The continuity of the eigenvalues suggests an expansion around
$q=\pi$, which leads to an analytical conjecture for the density of structures
in the $N$-dimensional binary space. 
Assuming the validity of this conjecture in more general models,
it provides a means of estimating density, and hence indirectly designability,
of structures by simply analyzing their sequences, without the need for
extensive enumerations of other possible structures.
 
The rest of the paper is organized as follows. In Section II we
review the hydrophobic model and the designabilities of structures. 
In Section III we discuss the methods and the results of PCA applied 
to the structure space, and relate the density and designability of
a structure to its projections onto the principal axes. 
In Section IV we demonstrate that Fourier transformation provides a
very good approximation to PCA, and show that in fact the two procedures
are equivalent for the subset of cyclic structures.
As a comparison with real structures, in Section V we introduce and study
an ensemble of pseudo-structures constructed by a Markovian process.
Finally, in Section VI we synthesize the numerical results of PCA analysis,
and develop a conjecture for the density of points in structure space.
 
\section{THE HYDROPHOBIC MODEL}

\begin{figure}
\vspace{4cm}
\centerline{\epsfxsize=8.5cm
\epsffile{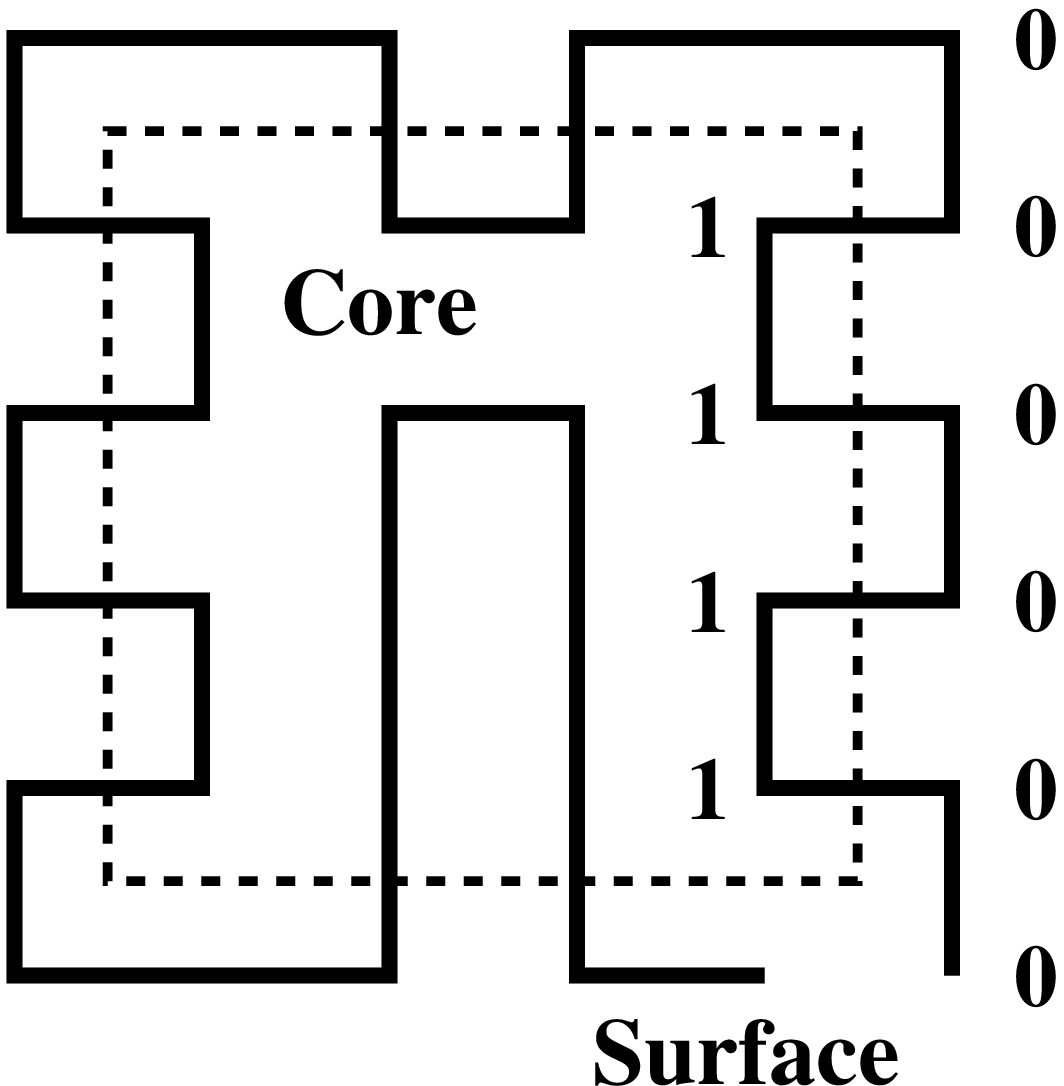}}                                        
\vspace {8cm}
\caption{Yahyanejad, et al.}
\label{lattice_protein}
\end{figure}

We start with a brief review of the hydrophobic model of Li {\em et al.} \cite{Li98}
and the designabilities of structures. 
Model sequences are composed of two types 
of amino acids, H and P. Each sequence $\{h_i\}$ (for $ i=1,2,\cdots,N$) is represented 
by a binary string or vector, with $h_i=0$ for a P-mer and $h_i=1$ for an H-mer. 
We take the polymer length $N=36$, for which there are $2^{36}$ sequences. Each
of these sequences can fold into any one of the many compact structures 
on a $6\times6$ square lattice (Fig.~\ref{lattice_protein}). There are 
$57,337$ such compact structures unrelated by rotation and mirror 
symmetries. In the hydrophobic model, the 
only contribution to the energy for a sequence folded into a structure is 
the burial of the H-mers in the sequence into the core of the structure. 
So if one represents a structure by a binary string or vector, $\{s_i\}$, for 
$i=1,2,\cdots,36$, with $s_i=0$ for the surface sites and $s_i=1$ for 
the core sites (Fig.~\ref{lattice_protein}), the energy is
\begin{equation}
E = - \sum_{i=1}^{N}{h_i s_i},
\label{ham}
\end{equation}
where $h_i$ is the sequence vector.

\begin{figure}
\vspace{6cm}
\centerline{\epsfxsize=8.5cm                                           
\epsffile{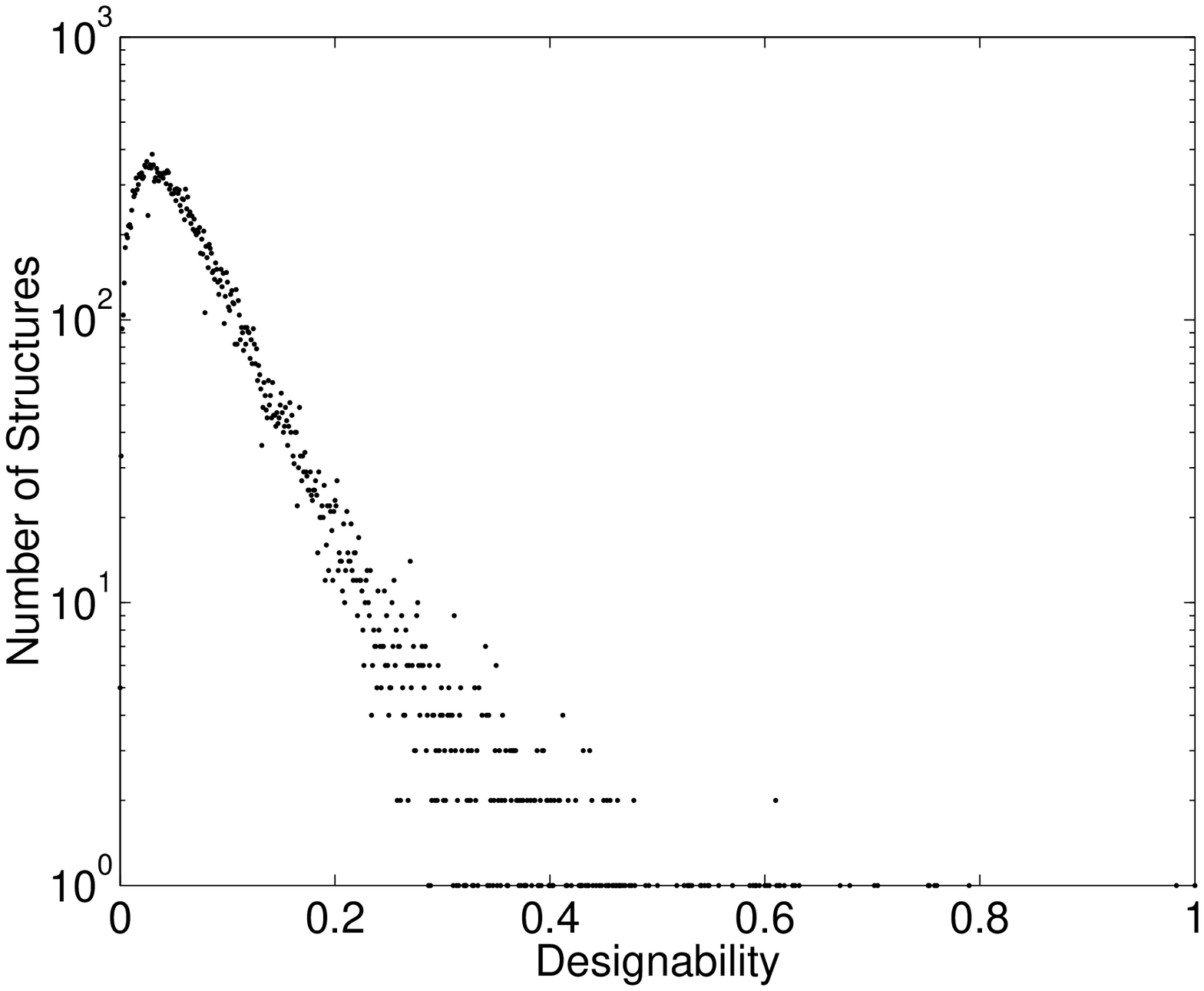}}
\vspace {8cm}
\caption{Yahyanejad, et al.}                                           
\label{hist_ds}
\end{figure}

The designability of a structure is defined as the number of sequences
that have the structure as their unique lowest-energy state. To obtain 
an estimate for designabilities of structures, we randomly sampled
$50,000,000$ sequences 
and found the unique
lowest-energy structure, if any, for each of them. 
In Fig.~\ref{hist_ds}, we plot the histogram of designabilities, {\em i.e.} number of 
structures with a given designability. Note that we have normalized 
designability so that its maximum value of $2981$ is scaled 
to one. In this paper, 
we define highly designable structures to be the top one percent of 
designable structures (structures with nonzero designability), which 
means $307$ structures with a designability larger than $0.47$. 

\begin{figure}
\vspace{6cm}
\centerline{\epsfxsize=8.5cm
\epsffile{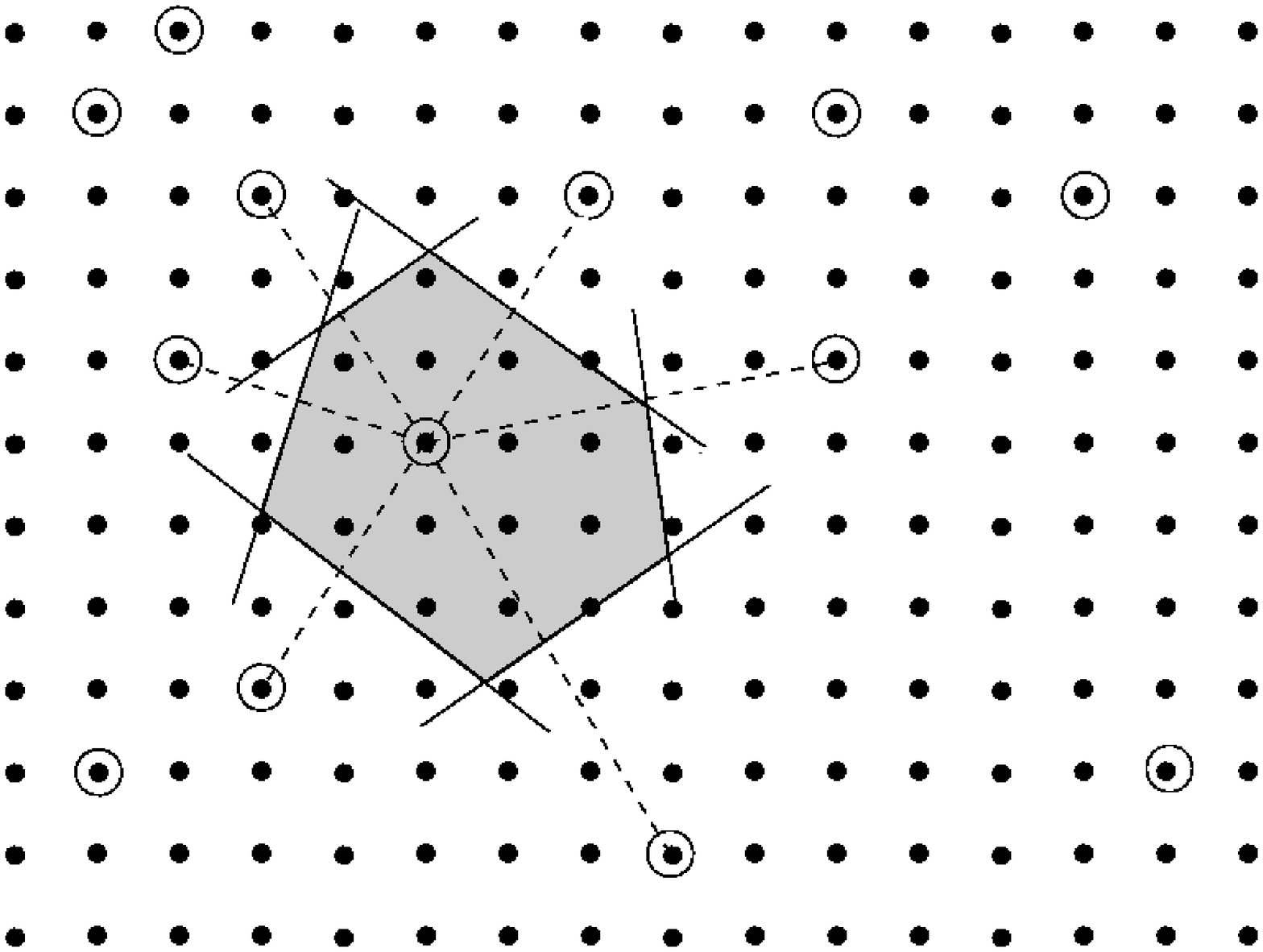}}
\vspace {8cm}
\caption{Yahyanejad, et al.}
\label{structure_space}
\end{figure}

In the hydrophobic model, both sequences and structures can be regarded
as points in a 36-dimensional binary space, or corners of a hypercube in
a Euclidean space of similiar dimension.
In this representation, the lowest-energy state of a sequence is simply
its nearest structure point. \cite{Li98}
Designabilities can then be obtained by constructing Voronoi polyhedra around
all points corresponding to structures in this space;
the designability of each structure is then the number of sequence points
that fall within the corresponding Voronoi polytope (Fig.~\ref{structure_space}). 
Structures in the lower density regions
have larger Voronoi polytopes and higher designability. 
Understanding how the structure points are distributed in 
this 36-dimensional space can thus help us address questions concerning 
designability. In the next section we examine the distribution 
of the structure points via the method of PCA. 

\section{PRINCIPAL COMPONENT ANALYSIS}

First, let us note that while sequences are uniformly distributed in the
36-dimensional hypercube, structures are distributed on a 34-dimensional
hyperplane because of the following two geometrical constraints. The
first constraint on structure vectors comes from the fact that all compact
structures have the same number of core sites, and thus
\begin{equation}
\sum_{i=1}^{36} s_i = 16.
\label{constraint1}
\end{equation}
The second constraint is that since the square lattice is bipartite, and any compact
structure traverses an equal number of `black' and `white' points, 
\begin{equation}
\sum_{i=1}^{36} (-1)^i s_i = 0.
\label{constraint2}
\end{equation}

Next, let us define the covariance matrix of the structure space as 
\begin{equation}
C_{i,j} = \left\langle s_i s_j\right\rangle - \left\langle s_i\right\rangle
\left\langle s_j\right\rangle , 
\end{equation}
where $i,j=1,2,\cdots,36$, and the average is 
over all the $57,337$ possible  $(s_1,s_2,\cdots,s_{36})$ for compact structures. 
The $36\times36$ covariance matrix is symmetric  $C_{i,j}=C_{j,i}$, and also
satisfies the condition $C_{i,j}=C_{37-i,37-j}$. The latter is due to
the reverse-labeling degeneracy of the structure ensemble, since if the
string $(s_1,s_2,\cdots,s_{36})$ is in this ensemble, then its
reverse $(s_{36},s_{35},\cdots,s_1)$ is also included.
This symmetry implies that if $(v_1,v_2,\cdots,v_{36})$ is an eigenvector
of the matrix $C_{i,j}$, then $(v_{36},v_{35},\cdots,v_1)$ is also an
eigenvector with the same eigenvalue. Therefore, for every eigenvector
of $C_{i,j}$ we have either $v_j=v_{37-j}$ or $v_j=-v_{37-j}$.

\begin{figure}                                                         
\vspace{6cm}                                                           
\centerline{\epsfxsize=8.5cm                                           
\epsffile{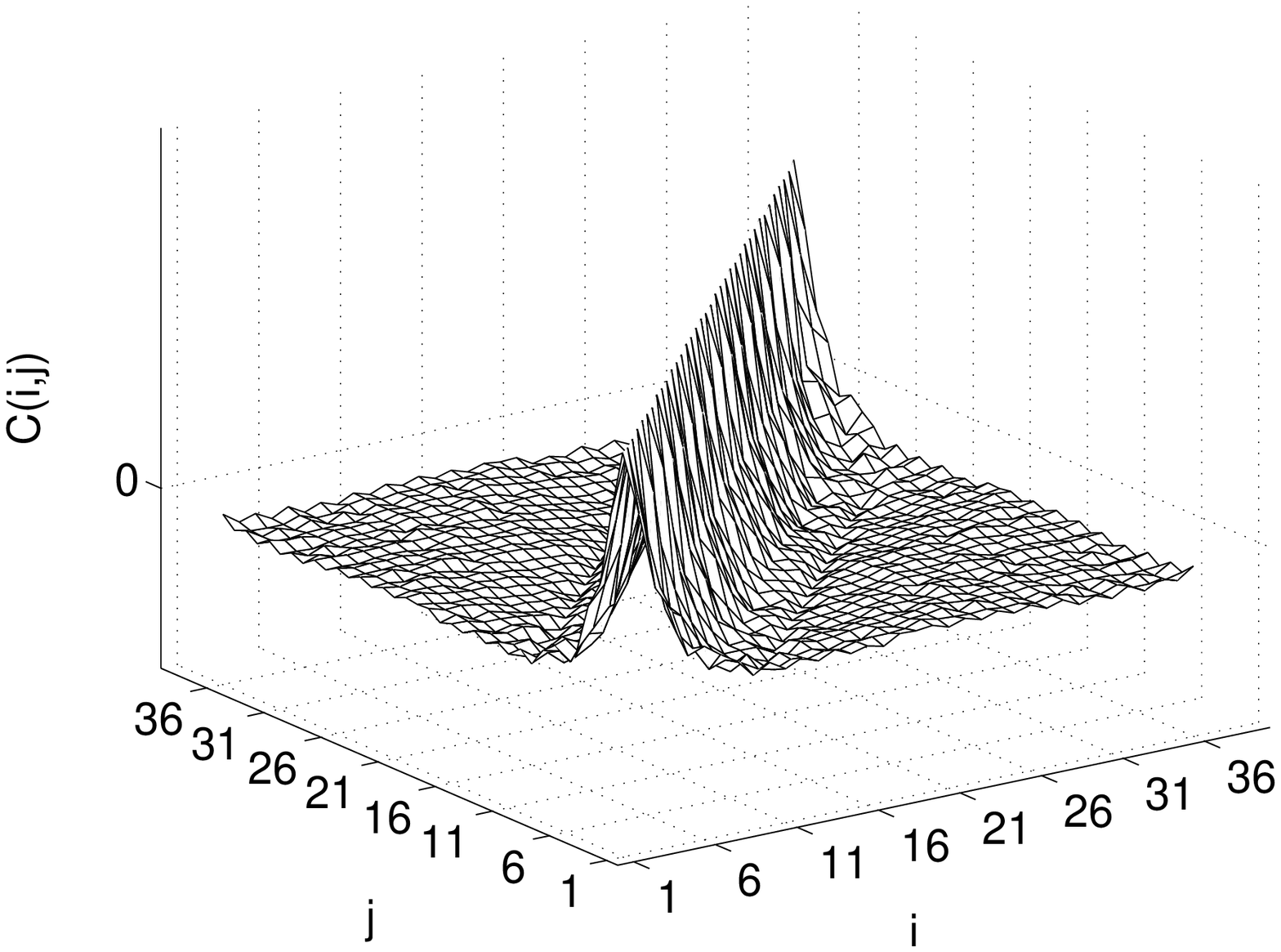}}                                             
\vspace {8cm}                                                          
\caption{Yahyanejad, et al.}                                           
\label{covariance}                                                     
\end{figure}

As depicted in Fig.~\ref{covariance}, the matrix
$C_{ij}$ is peaked along the diagonal and decays off-diagonally
with short range correlations. This feature reflects a general property of
compact self-avoiding walks; if a monomer is in the core (on the surface)
the neighboring monomers along the chain have enhanced probability to 
be in the core (on the surface). 
Another characteristic of $C_{ij}$ is that it is almost a function of $|i-j|$
only, {\em i.e.} $C_{ij}\approx F(|i-j|)$, barring some small end and 
parity effects. We expect this feature of approximate translational 
invariance to be generic beyond the $6\times6$ lattice model studied here.
We also looked at the covariance matrix for the 
subset of highly designable structures. 
While qualitatively similar, it tends to decay faster off-diagonally
than that of all structures.
This is attributed to the fact
that highly designable structures tend to have more frequent transitions
between core and surface sites. \cite{Li98,Cejtin02,Shih00}

\begin{figure}                                                         
\vspace{6cm}                                                           
\centerline{\epsfxsize=8.5cm                                           
\epsffile{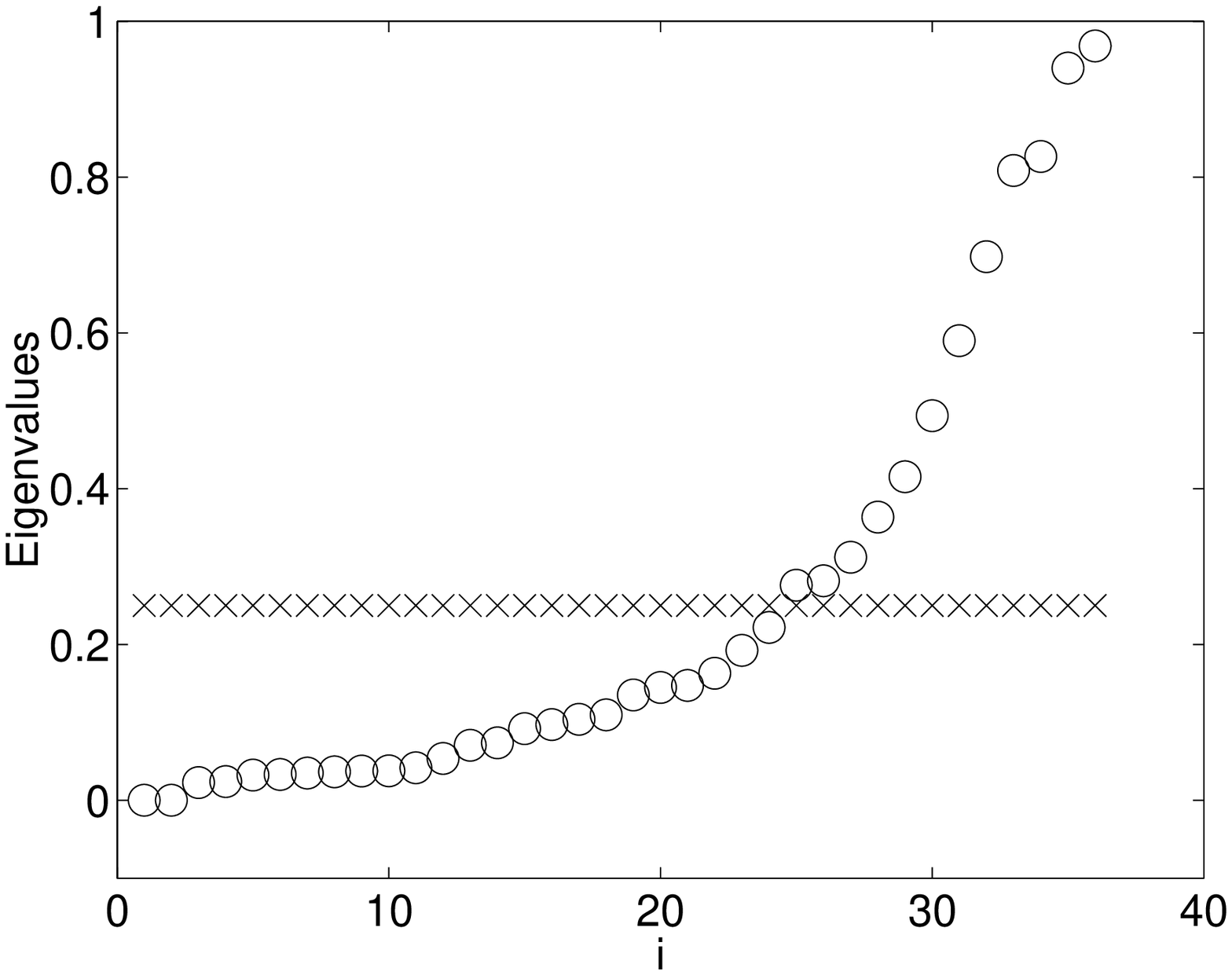}}                                            
\vspace {8cm}                                                          
\caption{Yahyanejad, et al.}                                           
\label{eigenvalues}                                                    
\end{figure}

For PCA of structure space, the matrix $C_{ij}$ is diagonalized to obtain its 
eigenvectors $\{\vec v\,^{(k)}\}$, and the corresponding eigenvalues $\{\lambda_k\}$ for
$k=1,2,\cdots,36$, which are shown in Fig.~\ref{eigenvalues}.
The two zero eigenvalues ($\lambda_1=\lambda_2=0$) result from the 
constraints in Eqs.~(\ref{constraint1}) and (\ref{constraint2}), with the
corresponding eigenvectors of $v_i^{(1)} = 1$, and 
$v_i^{(2)} = (-1)^i$ for $i=1,2,\cdots,36$, respectively. 
The remaining 34 nonzero eigenvalues range smoothly from zero
to one, making any further dimensional reduction not obvious.  
For comparison, the 36 eigenvalues of the uniformly distributed points of sequence
space are  all the same ($\lambda=1/4$). (It is easy to show that the 
covariance matrix for the sequence space is $C_{ij}=\delta_{ij}/4$.)
On the other hand, a uniform distribution on the 34-dimensional 
hyperplane where the structure points reside would result in 34 identical
eigenvalues of $360/1377 \approx 0.26$.\cite{footnote} 

\begin{figure}
\vspace{6cm}
\centerline{\epsfxsize=14cm
\epsffile{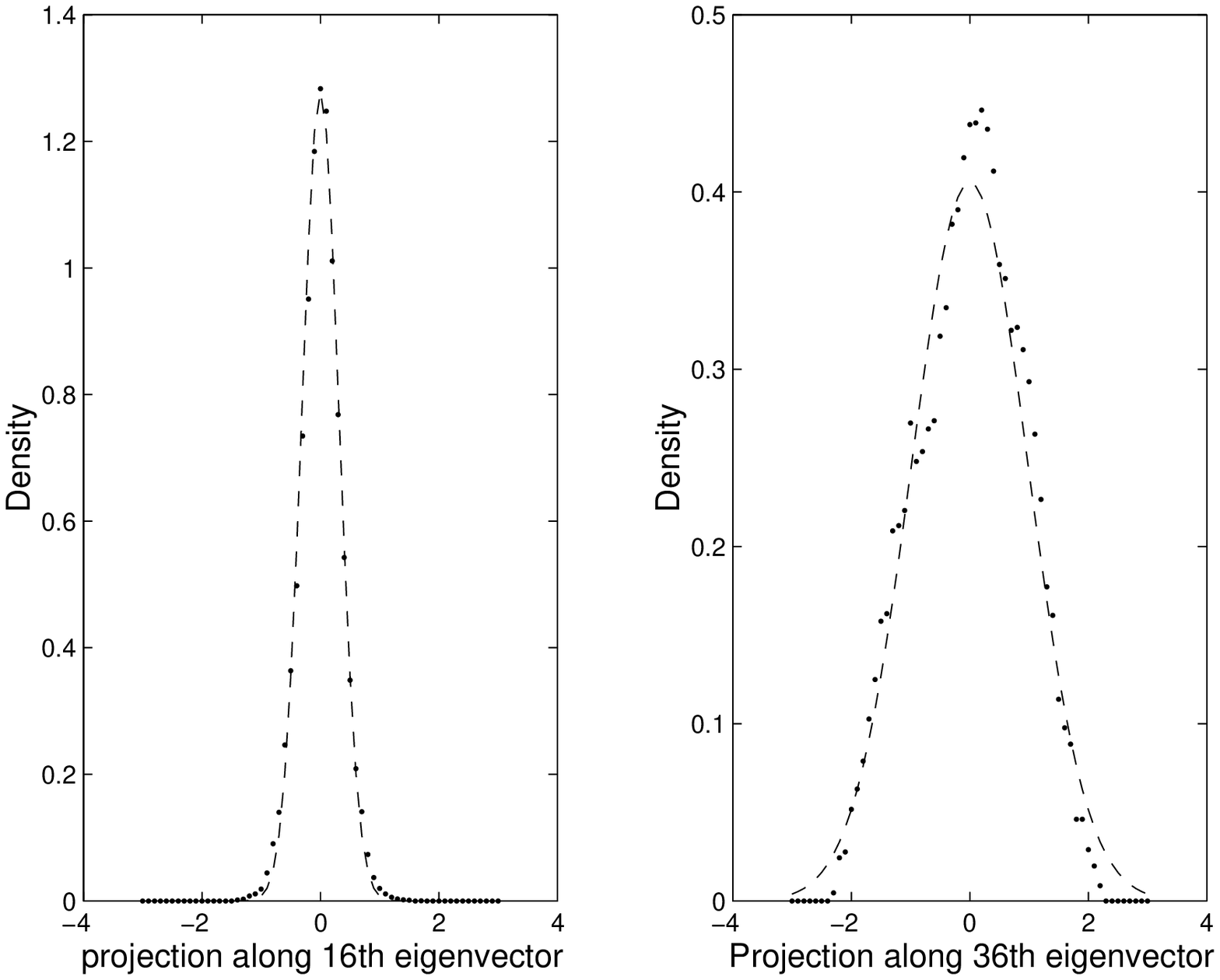}}
\vspace {8cm}
\caption{Yahyanejad, et al.}
\label{projections_gauss}
\end{figure}

Identification of the principal axes and eigenvalues does not necessarily
provide information about the distribution of points in space.
To examine the latter, we first project each structure vector onto its components
along the eigenvectors. Using the rotation matrix
$R_{ki}$ that diagonalizes the covariance matrix,
the component $y_k$ of the structure vector along principal axis $k$ is obtained as
\begin{equation}
y_k = \sum_{i=1}^{36} \left(s_i - \left\langle s_i\right\rangle \right) R_{ki}.
\end{equation}
Interestingly, we find that along each of the principal directions, the distribution
of components is a bell-shaped function with a single peak at zero.
Such distributions can then be well approximated by Gaussians whose
variances are the corresponding eigenvalues $\lambda_k$, {\em i.e.}
\begin{equation}
\rho_k(y_k) \approx\frac{1}{\sqrt{2\pi\lambda_k}} e^{- \frac{y_k^2}{2 \lambda_k}}.
\label{gauss}
\end{equation}
In Fig.~\ref{projections_gauss} we show the distribution of projections
$y_k$ on two principal axes $k=16$ and $k=36$, along with the 
corresponding Gaussian distributions.
 
\begin{figure}
\vspace{6cm}
\centerline{\epsfxsize=8.5cm
\epsffile{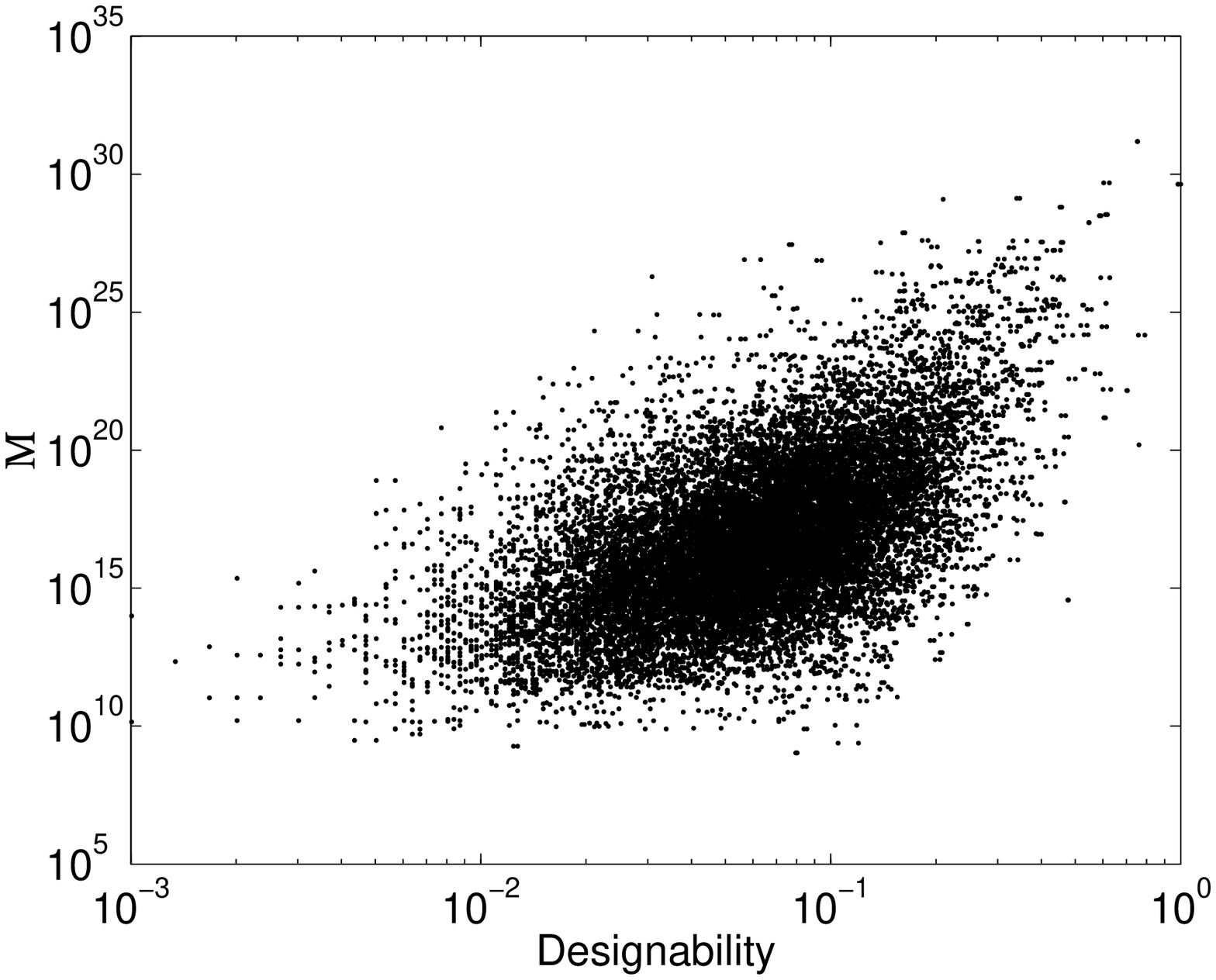}}
\vspace {8cm}
\caption{Yahyanejad, et al.}                                           
\label{correlation}
\end{figure}

Equation~(\ref{gauss}) provides a good characterization of the density of
structures in the $N$ dimensional space.
Highly designable structures are expected to lie in regions of this space
where the density of structures is small, while the number
of available sequences is large.
Let us consider a structure characterized by a vector $\vec y$. 
If the density of structural points in the vicinity of this point is 
$\rho_{str}({\vec y}\,)$, the number of available structures in a volume
$V$ around this point is $V\rho_{str}({\vec y}\,)$.
Neglecting various artifacts of discreteness, the volume of the Voronoi polyhedron
(see Fig.~3) around this point is given by $V({\vec y}\,)\approx1/\rho_{str}({\vec y}\,)$.
The designability is the number of structures within this volume, and
estimated as $\rho_{seq}({\vec y}\,)/\rho_{str}({\vec y}\,)$, where $\rho_{seq}({\vec y}\,)$
is the density of sequences. 
The sequence density is in fact uniform in the $N$-dimensional space.
The structure density can be approximated as the product of Guassians along
the principal projections, and thus 
\begin{equation}
 {\rm Designability}\approx \frac{\rho_{seq}({\vec y}\,)}{\rho_{str}\left({\vec y}\,\right)}
\propto \prod_{k=3}^{36} \frac{1}{\rho_k\left(y_k\right)}\propto
\exp\left[\sum_{k=3}^{36}\frac{y_k^2}{2 \lambda_k}\right]
\equiv {\cal{M}}({\vec y}\,).
\label{density_rel}
\end{equation}
We have neglected various proportionality constants in the above equation,
leading to the quantity ${\cal{M}}({\vec y}\,)$ which is our estimator for designability.
In Fig.~\ref{correlation},  the estimate $\cal{M}$ is plotted against the actual designability
for all designable structures. There is a reasonably good, but by no means
perfect, correlation between the designability and the estimator $\cal{M}$.
The structures with the top one percent value of $\cal{M}$ include 39\% of
the highly designable structures.

\section{FOURIER DECOMPOSITION  AND CYCLIC STRUCTURES}

In discussing Fig.~\ref{covariance}, we already noted that the covariance matrix
$C_{ij}$ is approximately a function of $|i-j|$, with corrections due to end effects.
If this were an exact symmetry, the matrix would be diagonal in the Fourier basis.
Even in the presence of the end effects, Fourier decomposition provides a very
good approximation to the eigenvectors and eigenvalues of PCA, as demonstrated below.
For each structure vector $\{s_j\}$, the Fourier components are obtained from
\begin{equation}
S_q = {1\over\sqrt{N}}\sum_{j=1}^N e^{iqj} \left(s_j-\left\langle s_j\right\rangle\right),
\label{sq}
\end{equation}
where $q=2\pi\alpha/N$, with $\alpha = 0,1,\cdots,N-1$. 
The average value of $\left\langle s_j\right\rangle$  
is subtracted for convenience.
With this subtraction, the  two constraints in Eqs.~(\ref{constraint1}) 
and (\ref{constraint2}) correspond to two zero modes in Fourier space,
as $S_0=0$ and $S_\pi=0$, and since $\{s_j\}$ are real $S^*_q = S_{-q}$.

The covariance matrix in the Fourier space is
\begin{equation}
\left\langle S_qS^*_{q'}\right\rangle = {1\over N}\sum_{j,j'=1}^N e^{i(qj-q'j')} C_{jj'},
\label{sqsq}
\end{equation}
and is both real and symmetric (since $C_{jj'}=C_{j'j}$).
If $C_{jj'}$ is translationally invariant, {\em i.e.} $C_{jj'}=F(|j-j'| \bmod N )$,
Eq.~(\ref{sqsq}) becomes
\begin{equation}
\left\langle S_qS^*_{q'}\right\rangle  = \delta_{q,q'} \lambda_q,
\label{q_mat}
\end{equation}
where 
\begin{equation}
\lambda_q = \sum_{k=0}^{N-1} e^{iqk} F(k) =\left\langle\left| S_q\right|^2\right\rangle,
\label{sq2}
\end{equation}
are the diagonal elements of the diagonalized matrix in Eq.~(\ref{q_mat}), and hence
the eigenvalues of $C_{jj'}$. Note that because the matrix is real--symmetric, its
eigenvalues  appear in pairs, {\em i.e.} 
\begin{equation}
\lambda_q=\lambda_{-q}\,.
\label{e-even}
\end{equation}

\begin{figure}                   
\vspace{6cm}                     
\centerline{\epsfxsize=8.5cm     
\epsffile{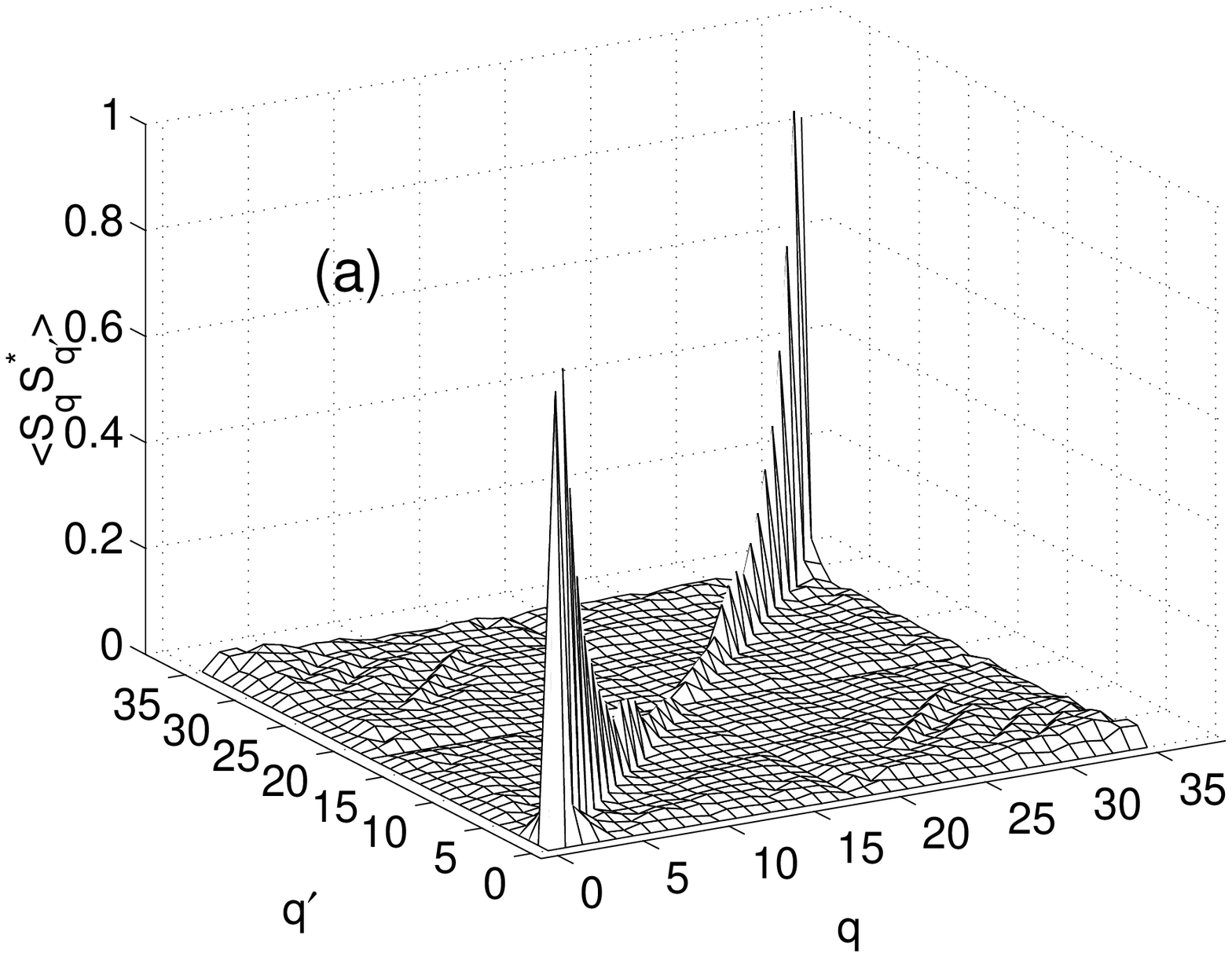}
\epsfxsize=8.5cm
\epsffile{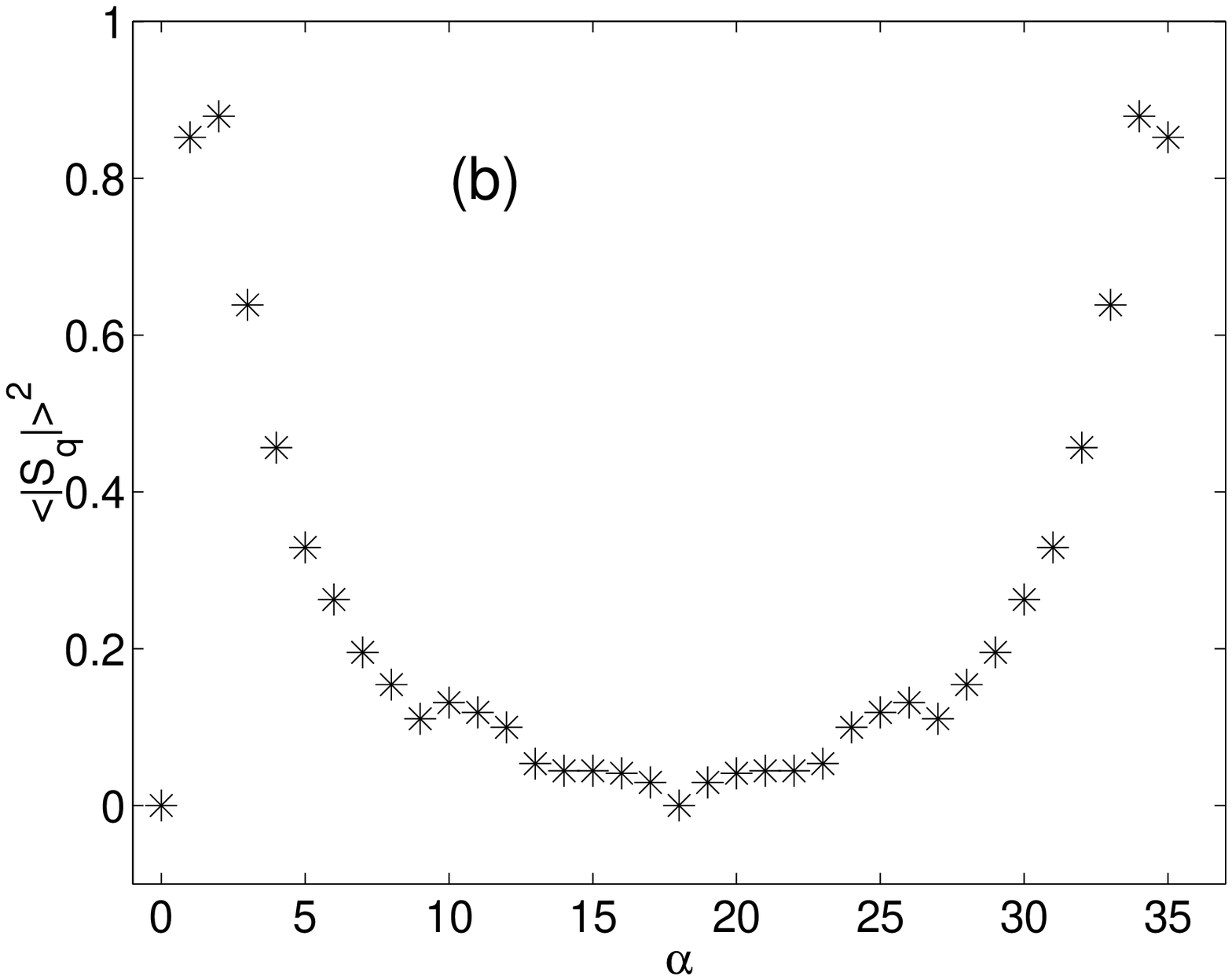}}   
\vspace {8cm}               
\caption{Yahyanejad, et al.}
\label{diag_f_exp_amp}
\end{figure}

\begin{figure}                   
\vspace{6cm}                     
\centerline{\epsfxsize=8.5cm     
\epsffile{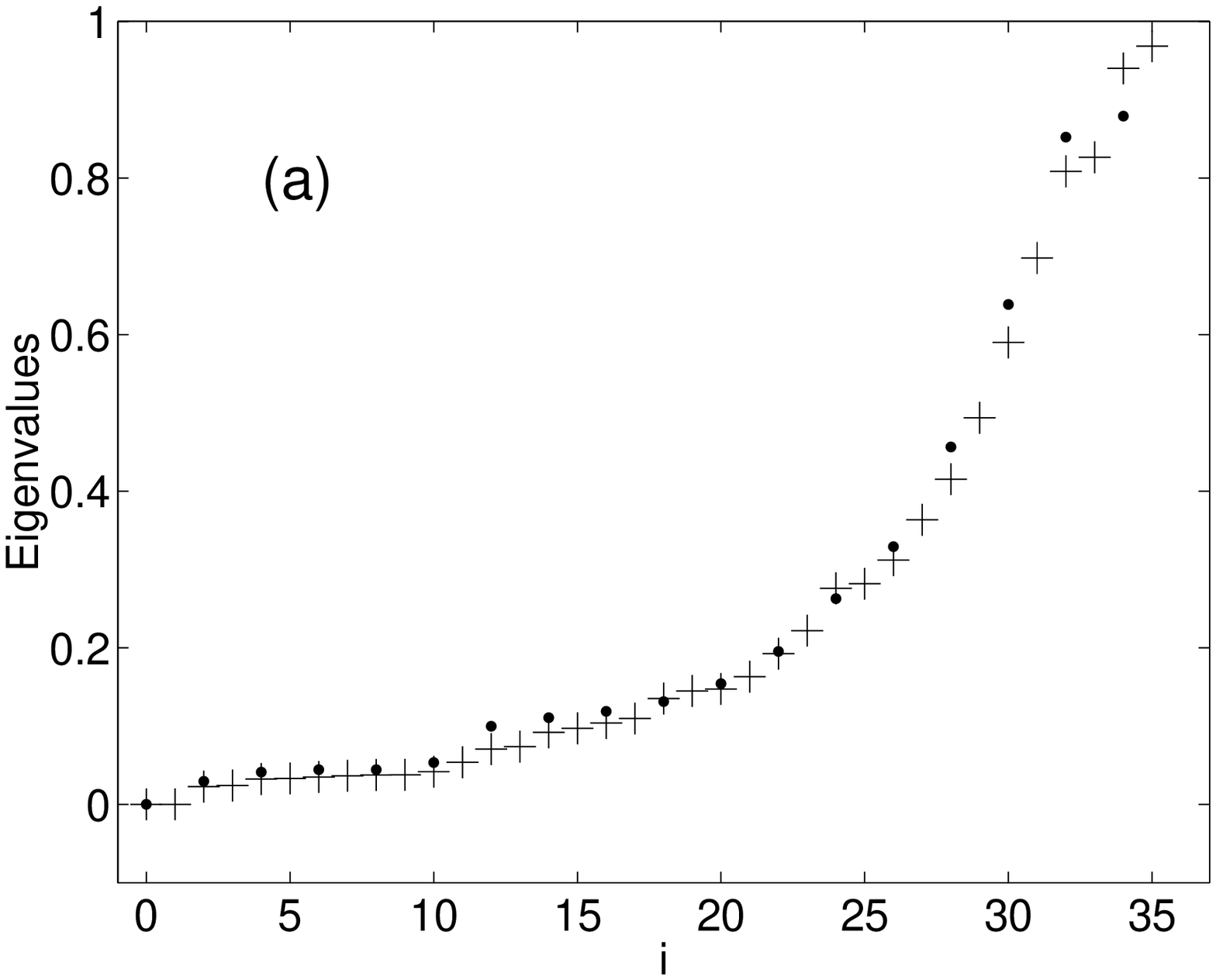}
\epsfxsize=8.5cm
\epsffile{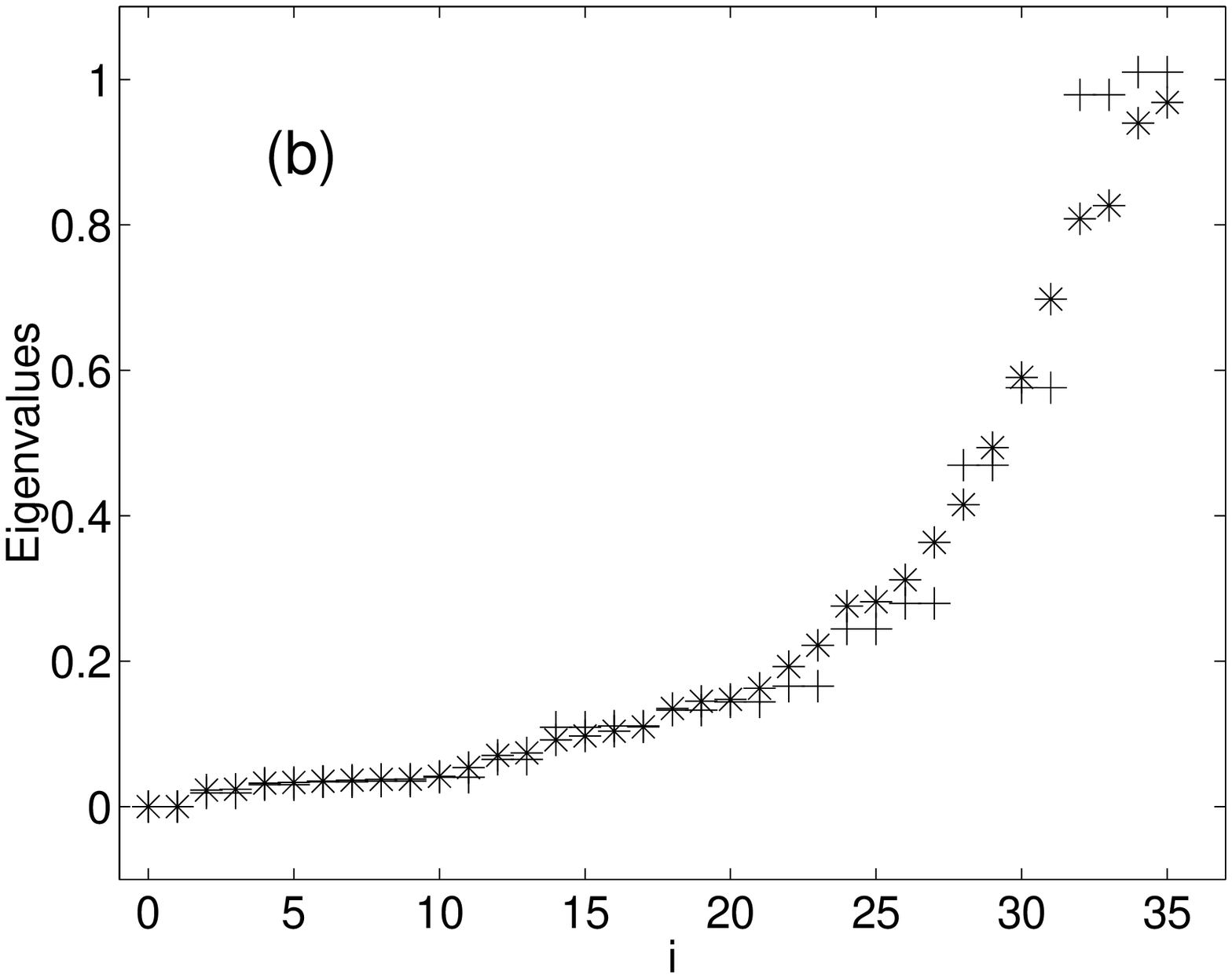}}
\vspace {8cm}
\caption{Yahyanejad, et al.}
\label{diag_eig}
\end{figure}

Since our covariance matrix is not fully translationally invariant, 
$\left\langle S_qS^*_{q'}\right\rangle$ is not diagonal.
However, as shown in Fig.~\ref{diag_f_exp_amp}a, its off diagonal elements are very small.
As required by symmetry, the diagonal elements form pairs of identical values. 
These diagonal elements, plotted versus the index $\alpha$ in 
Fig.~\ref{diag_f_exp_amp}b,
should provide a good approximation to the eigenvalues obtained by PCA.
This is corroborated in Fig.~\ref{diag_eig}a, where we compare 
$\left\langle|S_q|^2\right\rangle$ with the true eigenvalues of the covariance matrix
$C_{jj'}$.

Finally, we note that the end effects that mar the translational invariance
of the covariance matrix are absent in the subset of {\em cyclic structures}. 
Any structure whose two ends are neighboring points on the
lattice can be made cyclic by adding the missing bond.
Any one of the $N=36$ bonds on the resulting closed loop can be
broken to generate an element of the original set of structures,
and the corresponding structure strings are cyclic permutations of each other.
Thus, the covariance matrix $C_{cyclic}(j,j')$
of the set of all cyclic structures is translationally invariant. 
In our model of $6\times 6$ compact polymers, there are a total of 
$36\times276$ cyclic structures.
The Fourier transform of their covaraince matrix is diagonal as expected,
with diagonal elements depicted in Fig.~\ref{diag_eig}b. 
The corresponding Fourier eigenvalues
are quite close to the eigenvalues of the full matrix obtained in the PCA
(Fig.~\ref{diag_eig}b).
Thus the end effects do not significantly modify the correlations, and this
is specially true for the smallest eigenvalues which make the largest
contributions to the density in Eq.~\ref{density_rel}.

\section{A MARKOVIAN ENSEMBLE of PSEUDO--STRUCTURES}\label{markov}

The geometry of the lattice, and the requirement of compactness constrain
the allowed structure strings of zeros and ones in a non-trivial fashion.
In our estimation of designabilities so far, we have focused on the 
covariance matrix which carries information only about two point correlations
along these strings.
In principle, higher point correlations may also be important, and we may
ask to what extent the covariance matrix contains 
the information about the structures' designabilities? 
As a preliminary test, we performed a comparative study with an artificial
set of strings, not corresponding to real structures, but constructed to have
a covariance matrix similar to true structures on the $6\times6$ lattice. 

Specifically, we generated a set of random strings $\{\vec{t}\,\}$, of zeros
and ones of length 36, using a third order Markov process as follows. 
For each string, the first element $t_1$ is generated with probability 
$P(t_1=1)=\left\langle s_1\right\rangle$, where $\left\langle s_1\right\rangle$ is
the fraction of the true structure strings with $s_1=1$. 
The second element $t_2$ is generated according to a transition probability 
$P(t_1\to t_2)$ which is taken to be the ``conditional probability''
$P(s_2|s_1)$ extracted from the true structure strings. 
The third  point $t_3$ is generated according to a transition probability 
$P(t_1t_2\to t_3)$ which is the ``conditional probability''
$P(s_3|s_1s_2)$ extracted from the true structure strings. 
All the remaining points $t_j$, $j=4,5,\cdots,36$, are generated according to 
the transition probabilities $P(t_{j-3}t_{j-2}t_{j-1}\to t_j)$ equal to
the true ``conditional probabilities'' $P(s_j|s_{j-3}s_{j-2}s_{j-1})$
of actual structures.  
Sequences that do not satisfy the global constrains of Eqs.~(\ref{constraint1})
and (\ref{constraint2}) are thrown out.
For every Markov string generated, we also put its reverse in the pool, 
unless the string is its own reverse. 

\begin{figure}
\vspace{6cm}
\centerline{\epsfxsize=8.5cm     
\epsffile{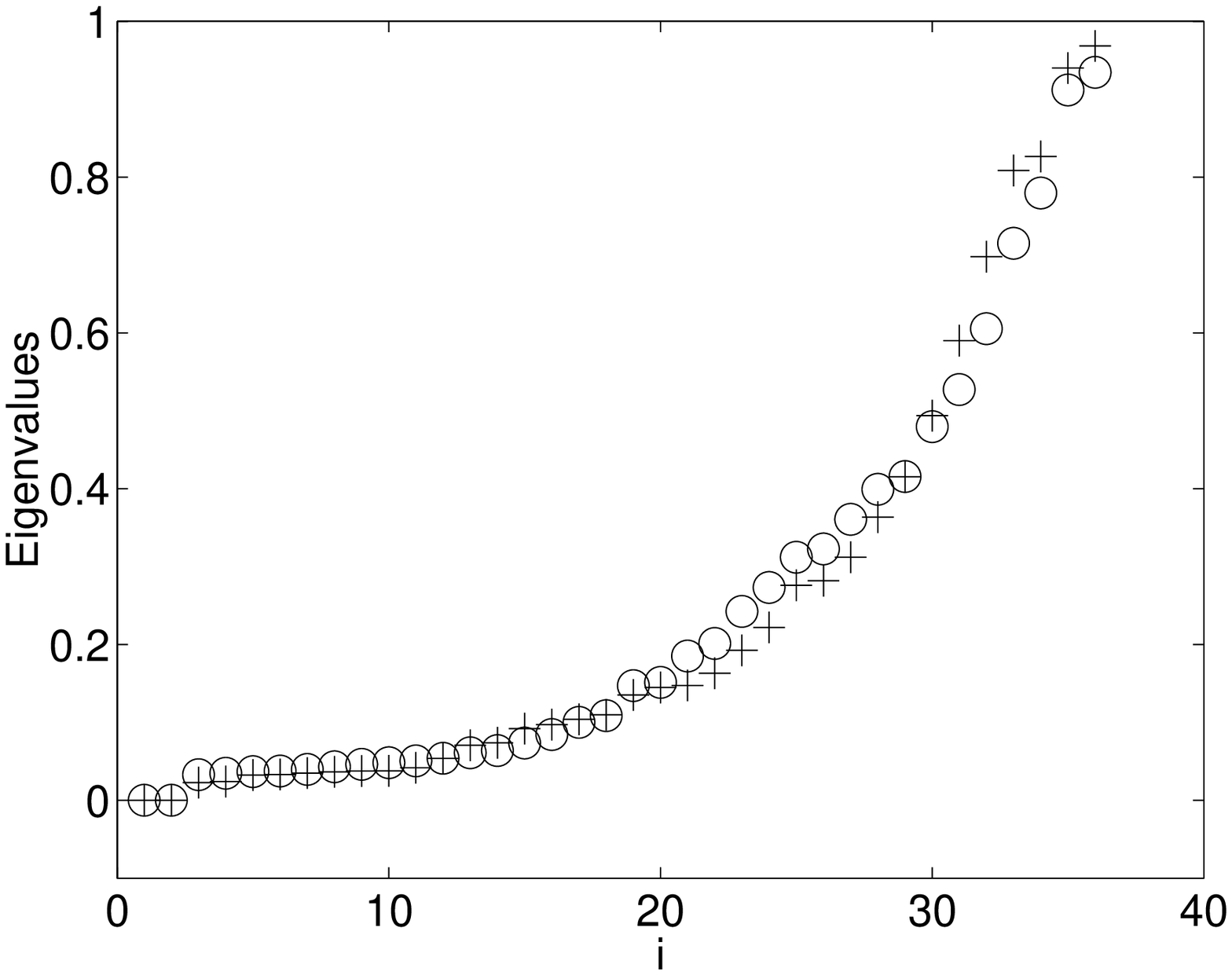}}
\vspace {8cm}                    
\caption{Yahyanejad, et al.}     
\label{eigenvalues_markov}        
\end{figure}

\begin{figure}
\vspace{6cm}
\centerline{\epsfxsize=8.5cm
\epsffile{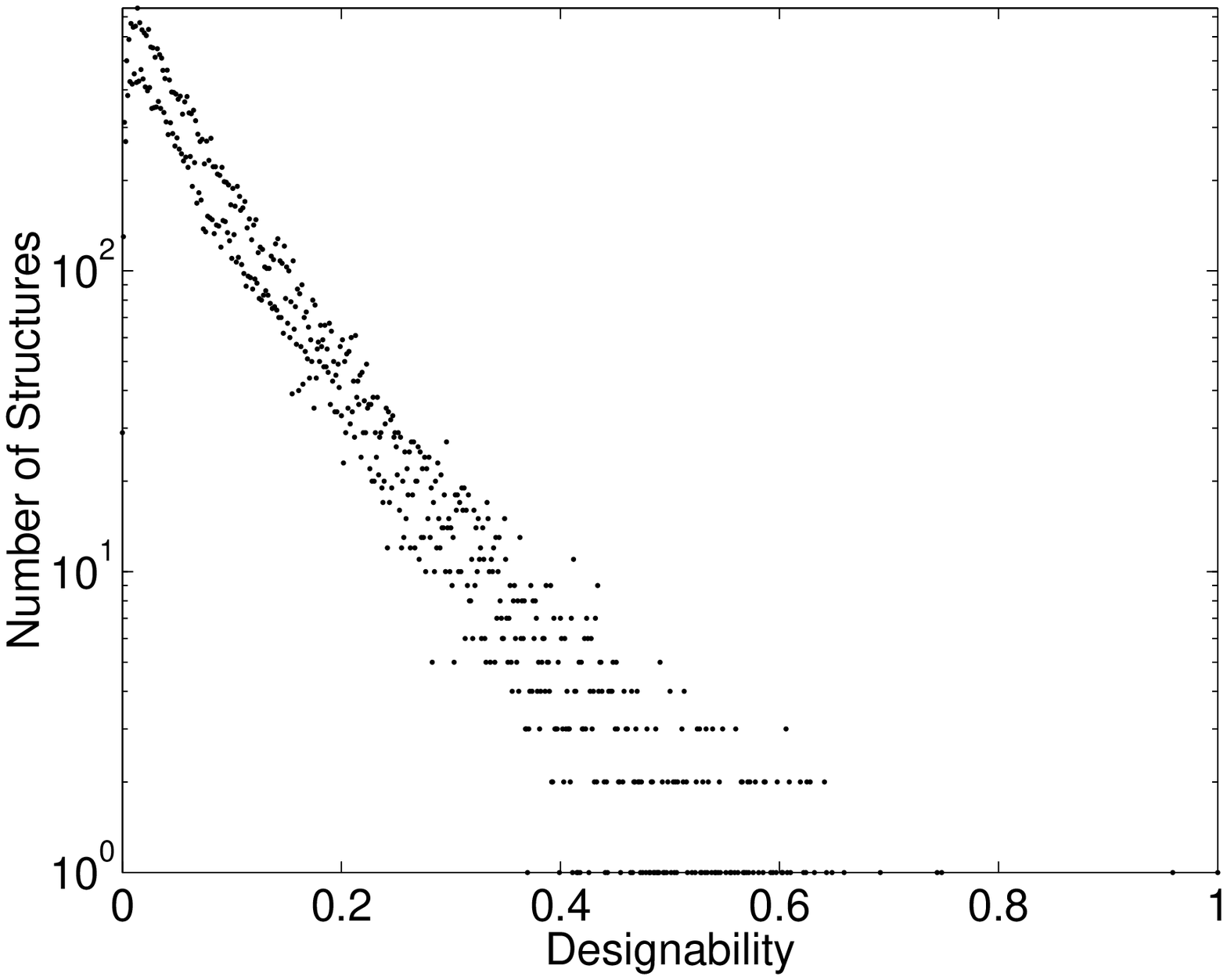}}
\vspace {8cm}
\caption{Yahyanejad, et al.}      
\label{hist_ds_markov}
\end{figure}

\begin{figure}
\vspace{6cm}
\centerline{\epsfxsize=8.5cm
\epsffile{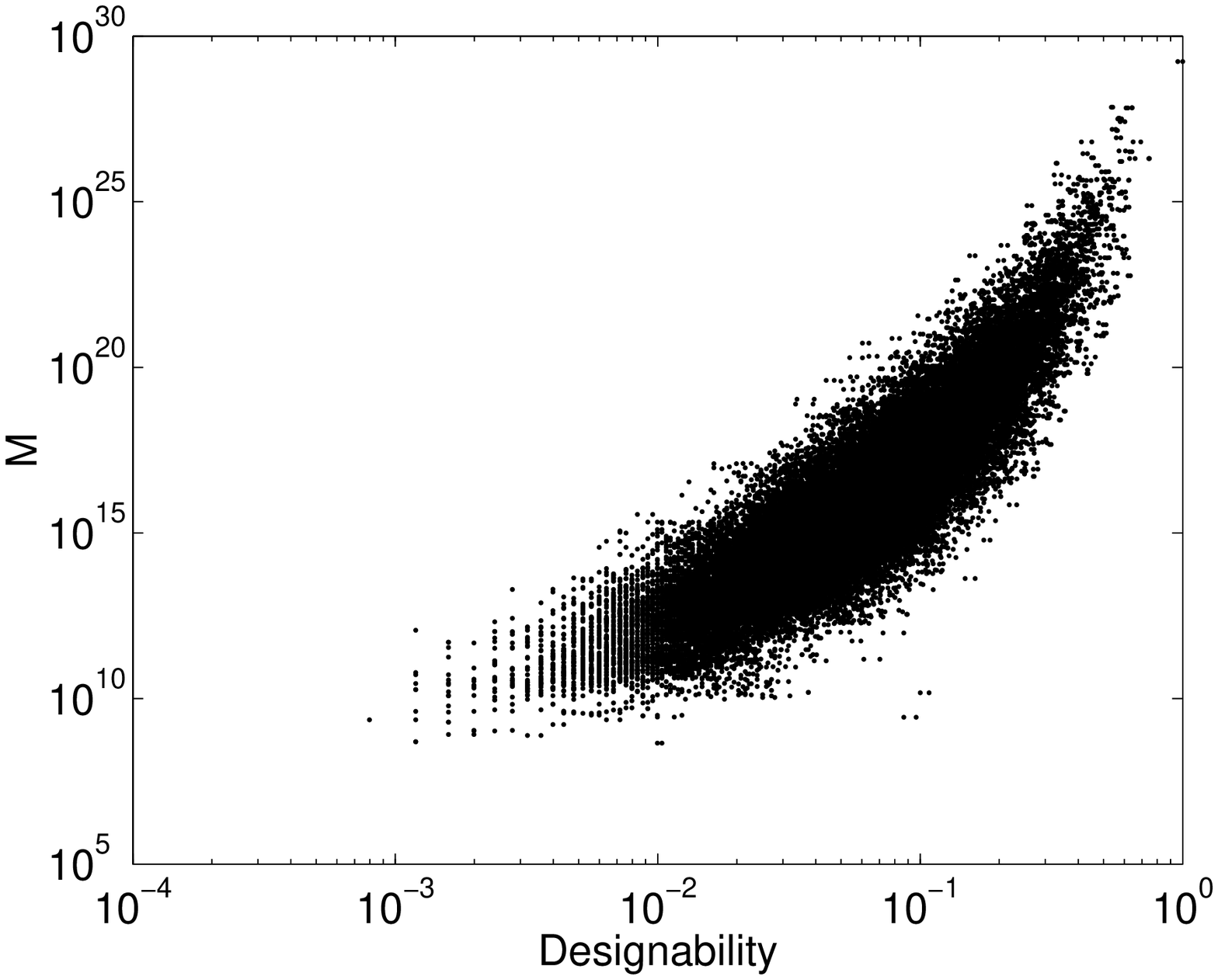}}
\vspace {8cm}
\caption{Yahyanejad, et al.}      
\label{correlation_markov}        
\end{figure}

The above Markovian ensemble has a covariance matrix, and corresponding
eigenvalues, very similar to those of the true structures,
as shown in Fig.~\ref{eigenvalues_markov}. 
We then calculated the designabilities of these ``pseudo-structures'' using
Eq.~(\ref{ham}) by uniformly sampling $5\times10^7$ random binary sequences.
The histogram of the designabilities (Fig.~\ref{hist_ds_markov}) is qualitatively 
similar to that of the true structures (Fig.~\ref{hist_ds}). 
Next we constructed the designability estimator $\cal{M}$
(Eq.~(\ref{density_rel})) for the pseudo-structures, using the
eigenvalues and eigenvectors of their covariance matrix. 
The quantity $\cal{M}$ is plotted versus designability in
Fig.~\ref{correlation_markov} for all the artificial
pseudo-structures with non-zero designability. The pseudo-structures with the top one percent
 value of $\cal{M}$ include 60\% of the highly designable psuedo-structures.

These results suggest that a considerable amount of information about 
the designability is indeed contained in the two point correlations of the string.
The designability estimator, Eq.~(\ref{density_rel}) in fact does a somewhat better job in
the case of pseudo-structures generated according to short-range Markov rules.

\section{CONCLUSIONS}

One of the most intriguing properties of compact structures, which emerged from
early extensive enumeration studies,\cite{Li96} is that designabilites range
over quite a broad distribution of values.
Such a large variation in designability is a consequence of a
non-uniform distribution of structure vectors, 
with highly designable structures typically found 
in regions of low density.\cite{Li98}
However, our study of  $6\times6$ lattice structures using PCA indicates that
the non-uniform density actually has a rather simple form that can be well
approximated by a single multi-variable Gaussian, as in Eq.~(\ref{density_rel}).
Since this method of estimating structure designability is based 
only on the overall distribution of structures, it can be a useful tool 
in cases where there is not enough computational power to enumerate the 
whole structure space and calculate the designability. 
To obtain an accurate enough covariance matrix requires only a uniform sampling 
of the structure space.

We can also attempt to use the numerical results as a stepping stone to
a more analytical approach for calculating the density of structures.
{\em First}, we note that the covariance matrix for all structures is rather similar to
that of the subset of cyclic structures, and that for the latter PCA is equivalent
to Fourier decomposition. 
{\em Second}, we observe that the multi-variable Gaussian approximation to
structure density in Eq.~(\ref{density_rel}) is most sensitive to the
eigenvalues that are close to zero.
In terms of Fourier components, these are eigenvalues corresponding to values
of $q$ close to $\pi$, and related to the constraint of Eq.~(\ref{constraint2}).
There is also a zero eigenvalue for $q=0$, related to condition~(\ref{constraint1}).
However, the latter global constraint appears not to have any local counterpart,
as there is a discontinuity in the eigenvalues close to zero. 
{\em Third},  the continuity of the eigenvalues as $q\to\pi$, along
with the symmetry of Eq.~(\ref{e-even}), suggests as expansion of the form
$\lambda_q=K(q-\pi)^2+{\cal O}\left((q-\pi)^4\right)$. 
Indeed the numerical results indicate that the important (smaller) eigenvalues 
can well be approximated by $K(q-\pi)^2$, with $K\approx 0.04$.\cite{fit}

With this approximation, the designability estimate of Eq.~(\ref{density_rel})
becomes
\begin{equation}
\label{Coulomb}
{\cal{M}}\left(\left\{\vec s\,\right\}\right)\approx
\exp\left[\sum_{q}\frac{\left|S_q\right|^2}{2 K(q-\pi)^2}\right]=
\exp\left[\frac{1}{2K}\sum_{i,j=1}^N (-1)^i s_i J_N\left(|i-j|\right) (-1)^j s_j\right].
\end{equation}
The first form in the above equation expresses the estimate in terms of the
Fourier modes of the structure string, while the second term is directly
in terms of the elements $\{s_i\}$.
The function $J_N\left(|i-j|\right)$ is the discrete Fourier transform of $1/q^2$,
which for large $N$ behaves as $|i-j|$.
Equation~(\ref{Coulomb}) is thus equivalent to the Boltzmann weight of
a set of unit charges on a discrete line of $N$ points marked by parity.
The charges on the sublattice of the same parity attract each other with a
potential $J_N(r)$, while those on different sublattices repel.
Such an interaction gives a larger weight (and hence designability) to
configurations in which the charges alternate between the core and surface
sites, as observed empirically.\cite{Li98,Cejtin02,Shih00}

In would be revealing to see how much of the above results, 
developed on the basis of a lattice hydrophobic model, can be applied to real
protein structures. One could use the exposure level of residues to the solvent 
in building up the structure vectors. 
Current methods deal with structure strings of a fixed length,
equal to the dimension of the structure space. 
Since real proteins have different lengths, there is a need for
a scaling method to handle them all together. 
Our study shows that the two point correlations of structure vectors
are approximately translationally invariant, and can be captured
by Fourier analysis.
This suggests the possibility of casting the density of points in
structure space in universal functional forms dependent only
of a few parameters encoding the properties of the underlying polymers.
If so, it would be possible to provide good estimates for designability
with polymers of varying length, without the need for extensive
numerical computations. 

\section{ACKNOWLEDGEMENT}

We thank G. Gatz, N. Butchler and E. Domany for useful discussions. 
The work was initiated, and partly completed, during the
Program on Statistical Physics and Biological Information at the
Institute for Theoretical Physics, UCSB, supported in part by the NSF
Contract No. PHY99-07949. We acknowledge support by the NSF Grant
No. DMR-01018213 (MY and MK), and an ITP Graduate Fellowship (MY).

\newpage                                                               
                                                                       
\begin{center}                                                         
FIGURE CAPTIONS                                                        
\end{center}                                                           
                                                                       
\begin{enumerate}                                                      

\item
A possible compact structure on the $6\times6$ square lattice. The 16 sites in the         
core region, enclosed by the dashed lines, are indicated by 1's;     
the 20 sites on the surface are labeled by 0's. Hence this structure
is represented by the string 001100110000110000110011000011111100. 
Note that each `undirected' open geometrical structure can be represented
by two `directed' strings, starting from its two possible
ends (except for structures with reverse-labeling symmetry where 
the two strings are identical).
It is also possible for the same string to represent different structures
which are folded differently in the core region.
For the $6\times6$ lattice of this study, there are
 $26929$ such `degenerate' structures, which are by definition non-designable.

\item
Number of structures with a given designability versus relative designability
for the $6\times6$ hydrophobic model. The data is generated by uniformly
sampling $5\times10^7$ strings from the sequence space. The
designability of each structure is normalized by the maximum possible
designability.

\item
Schematic representation of the 36-dimensional space in which sequences and 
structures are vectors or points. Sequences, represented by dots, are
uniformly distributed in this space. Structures, represented by circles,
occupy only a sparse subset of the binary points and are
distributed non-uniformly.
The sequences lying closer to a particular structure than to any other,
have that structure as their unique ground state. The designability of 
a structure is therefore the number of sequences lying entirely within 
the Voronoi polytope about that structure.

\item                                                                  
Covariance matrix $C_{ij}$ of all compact structures of the $6\times6$ square.

\item                                                                  
Eigenvalues of the covariance matrix for the structure vectors (circles), 
and for all points in sequence space (crosses). 

\item
Distributions of projections $y_k$ onto principal axes $k=16$ (a), and
$k=36$ (b), for all 57337 structures (dots). Also plotted are
Gaussian forms with variances $\lambda_{16}$ and $\lambda_{36}$, 
respectively (dashed lines). 


\item
The estimate $\cal{M}$ (Eq.~(\ref{density_rel})) versus scaled 
designability for all designable structures on the $6\times6$ square.

\item
(a) The Fourier transformed covariance matrix 
$\left\langle S_qS^*_{q'}\right\rangle  $ (Eq.~(\ref{sqsq})); and
(b) its diagonal elements $\left\langle |S_q|^2\right\rangle$.

\item
(a)The diagonal elements $\left\langle |S_q|^2\right\rangle $ (dots) plotted together with the
eigenvalues of the covariance matrix (pluses).
(b) Diagonal elements of the Fourier transformed $C_{cyclic}$, which are also the eigenvalues of the covariance matrix of cyclic structures (pluses).
Eigenvalues of the covariance matrix for all structures are indicated by stars.

\item
Eigenvalues of the covariance matrix for the structures generated
by the Markov model (circles), and that of the true structure space
(pluses).

\item
Number of pseudo-structures with a given designability versus designability 
for the pseudo-structure strings randomly generated using the Markov model. 
The data is generated by uniformly sampling $5\times10^7$ binary sequence 
strings.
The designability of each pseudo-structure is normalized by the maximum possible
designability.

\item
The quantity $\cal{M}$ versus designability for all designable
pseudo-structures generated by the Markov model. 
 
\end{enumerate}

\newpage 

\newpage

\newpage

\newpage                                                               

\newpage                                                               

\newpage

\newpage

\newpage                         

\newpage                         

\newpage

\newpage

\newpage


\newpage

\begin{references}                                                     

\bibitem{Finkel87}
A. V. Finkelstein and O. B. Ptitsyn, Prog. Biophys. Mol. Biol.
{\bf 50}, 171 (1987).
                                                                       
\bibitem{Chothia92}
C. Chothia, Nature {\bf 357}, 543 (1992).
                                                                       
\bibitem{Brenner97}
S. E. Brenner, C. Chothia, and T. J. P. Hubbard, Curr. Opin. in
Struct. Biol. {\bf 7}, 369 (1997).
                                                                       
\bibitem{Orengo94}
C. A. Orengo, D. T. Jones, and J. M. Thornton,
Nature {\bf 372}, 631 (1994).
                                                                       
\bibitem{Wang96}
Z. X. Wang, Proteins {\bf 26}, 186 (1996).
                                                                       
\bibitem{Govin99}
S. Govindarajan, R. Recabarren, and R. A. Goldstein,
Proteins {\bf 35}, 408 (1999).
                                                                       
\bibitem{Camacho93}
C. J. Camacho and D. Thirumalai, Phys. Rev. Lett. {\bf 71}, 2505
(1993).
                                                                       
\bibitem{Yue95}
K. Yue and K. A. Dill,
Proc. Natl. Acad. Sci. USA {\bf 92}, 146 (1995).
                                                                       
\bibitem{Govin95}
S. Govindarajan and R. A. Goldstein,
Biopolymers {\bf 36}, 43 (1995).
                                                                       
\bibitem{Li96}
H. Li, R. Helling, C. Tang, and N. S. Wingreen,
Science {\bf 273}, 666 (1996).
                                                                       
\bibitem{Govin96}
S. Govindarajan and R. A. Goldstein,
Proc. Natl. Acad. Sci. USA {\bf 93}, 3341 (1996).
                                                                       
\bibitem{Li98}
H. Li, C. Tang, and N. S. Wingreen,
Proc. Natl. Acad. Sci. USA {\bf 95}, 4987 (1998).
                                                                       
\bibitem{Buchler00}
N. E. G. Buchler and R. A. Goldstein,
J. Chem. Phys. {\bf 112}, 2533 (2000).
                                                                       
\bibitem{Helling01}
R. Helling, H. Li, J. Miller, R. M\'elin, N. Wingreen, C. Zeng, and C.
Tang, J. Mol. Graph. Model. {\bf 19}, 157 (2001).
                                                                       
\bibitem{Cejtin02}
H. Cejtin, J. Edler, A. Gottlieb, R. Helling, H. Li, J. Philbin,
N. Wingreen, and C. Tang, J. Chem. Phys. {\bf 116}, 352 (2002).
                                                                       
\bibitem{Miller02}
J. Miller, C. Zeng, N.S. Wingreen, and C. Tang, Proteins {\bf 47}, 506
(2002).
                                                                       
\bibitem{Melin99}
R. M\'elin, H. Li, N. Wingreen, and C. Tang, J. Chem. Phys. {\bf 110},
1252 (1999).
                                                                       
\bibitem{Wang00}
T. Wang, J. Miller, N. Wingreen, C. Tang, and K. Dill, J. Chem. Phys.
{\bf 113}, 8329 (2000).
                                                                       
\bibitem{Finkel93}
A. V. Finkelstein, A. M. Gutin, and A. Y. Badretdinov,
FEBS Lett. {\bf 325}, 23 (1993).
                                                                       
\bibitem{Finkel95}
A. V. Finkelstein, A.  Y. Badretdinov, and A. M. Gutin, 
Proteins  {\bf 23}, 142 (1995).
                                                                       
\bibitem{Govin97}
S. Govindarajan and R. A. Goldstein,
Biopol. {\bf 42}, 427 (1997).

\bibitem{Shih00}
C. T. Shih, Z. Y. Su, J. F. Gwan, B. L.  Hao, C. H. Hsieh, and H. C. and Lee,
Phys. Rev. Lett. {\bf 84}, 386 (2000).

\bibitem{footnote}
This can be seen from the following argument:
Since all points in the 34-dimensional hyperplane are equivalent
up to  parity, the most general form of the covariance matrix is
$C_{ij}=x+y(-1)^{i-j}+z\delta_{ij}$. Requiring zero eigenvalues for
eigenvectors $(1,1,1,1,...)$ and $(-1,1,-1,1,...)$ gives the constraints
$36x+z=0$, and $36y+z=0$, i.e. $x=y=-z/36$. The value of $z$ is then set
by $C_{ii}=\left\langle s_i\right\rangle  (1-\left\langle s_i\right\rangle  )=20/81$, where $\left\langle s_i\right\rangle  =16/36$. So we have
$x=y=-10/1377$ and $z=360/1377$. It is then easy to see that the 34
nonzero eigenvalues are $360/1377$.

\bibitem{fit}
If  not forced to go through zero for $q=\pi$, a somewhat better fit can be obtained
with $\lambda_q=0.03+0.04(q-\pi)^2+{\cal O}\left((q-\pi)^4\right)$.
Fourier transforming back to real space, the additional constant leads to screened
Coulomb interactions for $J_N(s)$. 

\end{references}
\end{document}